\documentclass[reprint,showpacs,amsmath,amssymb]{revtex4-2}
\usepackage{graphicx}% Include figure files
\usepackage{natbib}
\usepackage{blindtext}
\usepackage{dcolumn}% Align table columns on decimal point
\usepackage{ulem}
\usepackage{bm}% bold math
\usepackage{amsmath}
\usepackage[utf8]{inputenc}
\usepackage{hyperref}
\usepackage{balance}
\usepackage{multirow}
\usepackage{longtable}
\usepackage{epstopdf}
\usepackage{comment}
\usepackage{xcolor}
\newcommand{\rrt}{\textcolor{red}}
\newcommand{\bbt}{\textcolor{blue}}
\usepackage{dcolumn}% Align table columns on decimal point
\renewcommand{\k}{{\bbox k}}
\renewcommand{\k}{{\bm k}}
\newcommand{\hs}{\hspace*}
\newcommand{\vs}{\vspace*}
\newcommand{\np}{\newpage}
\newcommand{\cx}{C$_{60}$~}
\newcommand{\eref}[1] {(\ref{#1})}
\newcommand{\Eref}[1] {Eq.~(\ref{#1})}
\newcommand{\Fref}[1] {Fig. \ref{#1}}
\newcommand{\Tref}[1] {Table \ref{#1}}
\newcommand{\ra}{\rangle}
\newcommand{\la}{\langle}
\newcommand{\nn}{\nonumber}
\newcommand{\be}{\begin{equation}}
\newcommand{\ee}{\end{equation}}
\newcommand{\br}{\begin{eqnarray*}}
\newcommand{\er}{\end{eqnarray*}}
\newcommand{\ba}{\begin{eqnarray}}
\newcommand{\ea}{\end{eqnarray}}
\newcommand{\bp}{\begin{minipage}}
\newcommand{\ep}{\end{minipage}}
\newcommand{\ds}{\displaystyle}
\newcommand{\bs}{\bigskip}
\newcommand{\bt}{\begin{tabular}}
\newcommand{\et}{\end{tabular}}
  \newcommand{\mc}{\multicolumn}
\renewcommand{\u}{\underline}
\renewcommand{\r}{{\bbox r}}
\renewcommand{\r}{{\bm r}}
\renewcommand{\l}{\lambda}
\newcommand{\n}{{\bbox n}}
\renewcommand{\n}{{\bm n}}
\newcommand{\e}{{\bbox e}}
\newcommand{\z}{{\bbox z}}
\renewcommand{\z}{{\bm z}}
\newcommand{\tri}[3]
{\left(\begin{array}{rrr}
{#1}&{#2}&{#3}\\
0&0&0\\
\end{array}\right)}

\newcommand{\three}[6]
{\left(\!\!\!\begin{array}{rrr}
{#1\!}&{#2\!}&{#3\!}\\
{#4\!}&{#5\!}&{#6\!}\\
\end{array}\!\right)}
\begin{document}

%\title{Zeptosecond dynamics in atoms: fact or fiction?}
\title{Zeptosecond dynamics in atoms: fact or fiction?}
%\title{Wigner smith time delay mechanism leading to nuclear phenomena being observed in atomic measurements}
%\title{Nuclear orbiting resonances in atomic phenomena}
\author{T. Nandi$^{1}$, Prashant Sharma$^2$, Soumya Chatterjee$^3$, D. Mitra$^3$, Adya P Mishra$^4$, Y. Azuma $^5$, F. Koike$^6$ and A.S. Kheifets$^7$} %V. S. Yakovlev$^5$,}
\affiliation{$^{1}$Inter-University Accelerator Centre, Aruna Asaf Ali Marg, Near Vasant Kunj, New Delhi-110067, India.}
\thanks {Email:\hspace{0.0cm} nanditapan@gmail.com. Present address: 1003 Regal, Mapsko Royal Ville, Sector-82, Gurgaon-122004, India.}
\affiliation{$^2$Department of Particle and Astrophysics, Weizmann Institute of Science, Rehovot 76100, Israel}
\affiliation{$^3$Department of Physics, University of Kalyani, Kalyani, Nadia-741235, WB, India. }
\affiliation{$^4$Atomic \& Molecular Physics Division, Bhabha Atomic Research Centre, Trombay, Mumbai - 400 085, India}
\affiliation{$^5$Deapartment of Physics, Indian Institute of Technology, New Delhi, Delhi-110016, India.}
\affiliation{$^6$ Sophia University, 7-1 Kioichō, Chiyoda City, Tokyo 102-8554, Japan}
\affiliation{$^7$Research School of Physics and Engineering, The Australian National University, Canberra,  ACT 0200, Australia  }
%\newpage
\begin{abstract}
Photon exchange due to nuclear bremsstrahlung during nuclear collisions can cause Coulomb excitation in the projectile and the target nuclei. The corresponding process originated in nuclear timescales can also be observed in atomic phenomenon experimentally if it delayed by at least with an attosecond or longer timescales. We have found that this happens due to a mechanism involving the Eisenbud-Wigner-Smith time delay process. We have estimated photoionization time delays in atomic collisions utilizing the nonrelativistic version of random phase approximation with exchange and Hartree-Fock methods. We present three representative processes in which we can observe the phenomena in attosecond timescales even though they originate from excitations in the zeptosecond timescales. Thus the work represents an investigation of parallels between two neighboring areas of physics. Furthermore the present work suggests new possibilities for atomic physics research near the Coulomb barrier energy, where the laser is replaced by nuclear bremsstrahlung.
\end{abstract}

%\pacs{25.70.Bc, 25.70.-z, 25.70.Lm, 34.50.Fa}

\maketitle
During inelastic collisions of charged projectiles with atoms, atomic Coulomb and Pauli excitation \cite{basbas1974universal} cause inner-shell vacancy production through direct ionization to the continuum of the target atoms or by electron capture from the target atoms into an unoccupied state of the projectile ions. This process, discovered in 1930s, is known as Coulomb ionization of inner shells by heavy charged particles \cite{brandt1974binding,brandt1979shell}. It is a typical atomic or electromagnetic process having significantly large difference in range and coupling constant from strong force and thus nuclear phenomena usually do not have any influence of the atomic processes. However, in early fifties, an experiment detects K x-rays accompanying  $\alpha$ decay process from radioactive $^{210}Po$ \cite{barber1952evidence} and this x-ray emission phenomena was not described by only the Coulomb ionization process \cite{ciocchetti1965k}. It suggested a coincidence experiment to observe it more discerningly.\\
%%%%%%%%%%%%%%%%%%%%%%%%%%%%%
\indent In late seventies, Blair et al. \citep{blair1978nuclear} succeeded in observing it for the first time through a clear rise in Ni K x-ray production cross-section measured in coincidence with elastically scattered protons, while its energy was passed over the s$_{1/2}$ nuclear resonance at 3.151 MeV. Subsequently, a few more such experimental evidences on enhanced K-shell ionization were found due to the s$_{1/2}$ nuclear resonance at 461 keV of $^{12}C$ \cite{duinker1980experimental} and at 5.060 MeV of $^{88}Sr$ \cite{chemin1981measurement}. A theoretical study \cite{blair1982theory} suggested that monopole excitation might be responsible in exhibiting the influence of nuclear resonance on atomic collision provided the resonance width was less than or equal to the K- shell binding energy or equivalently, the time delay must be $\geq$ the K-shell orbiting period. On the other side, Greenberg et al. \cite{greenberg1977impact} observed large enhancement of K-shell ionization cross-section at small impact parameters in a heavy-ion collision experiment at energies above the Coulomb barrier. This fact was interpreted as a contribution of nuclear rotational-coupling mechanism  in addition to radial coupling proposed by Betz {\it et. al.} \cite{betz1976direct}. \\
%%%%%%%%%%%%%%%%%%%%%%%%%%%%%%%%%%%%%%%%%%%%%%%%%%%%%%%%
\indent About 40 years later, Sharma and Nandi \cite{Prashant-PRL-2017} have observed unusual resonance-like structures in the K x-ray spectra as the beam energy tending to the fusion barrier energy. The resonance structures were observed in the K x-ray energy of the elastically scattered projectile ion spectrum versus the beam energy plot \cite{Prashant-PRL-2017}. Such resonances have been attributed to the shakeoff ionization due to sudden nuclear recoil. Note that projectile x-ray energy corresponds to the mean charge state of the projectile ions inside the target foil and the higher x-ray energy implies the higher charge state \cite{Prashant-PLA}. Thus, variation of K x-ray energy is nothing but the variation of mean charge state of the projectile ions with beam energy. Besides the nuclear recoil, another important process called nuclear bremsstrahlung  \cite{jakubassa1975bremsstrahlung} is prevalent, which causes the above-mentioned nuclear resonances \cite{griffy1962nuclear}, nuclear Coulomb excitation \cite{Alder}, nuclear giant dipole resonances \cite{izumoto1981coulomb}, etc. \\
%%%%%%%%%%%%%%%%%%%%%
\begin{figure}[!h]
  \centering
   \includegraphics[width=8.5cm, height=4.0cm]{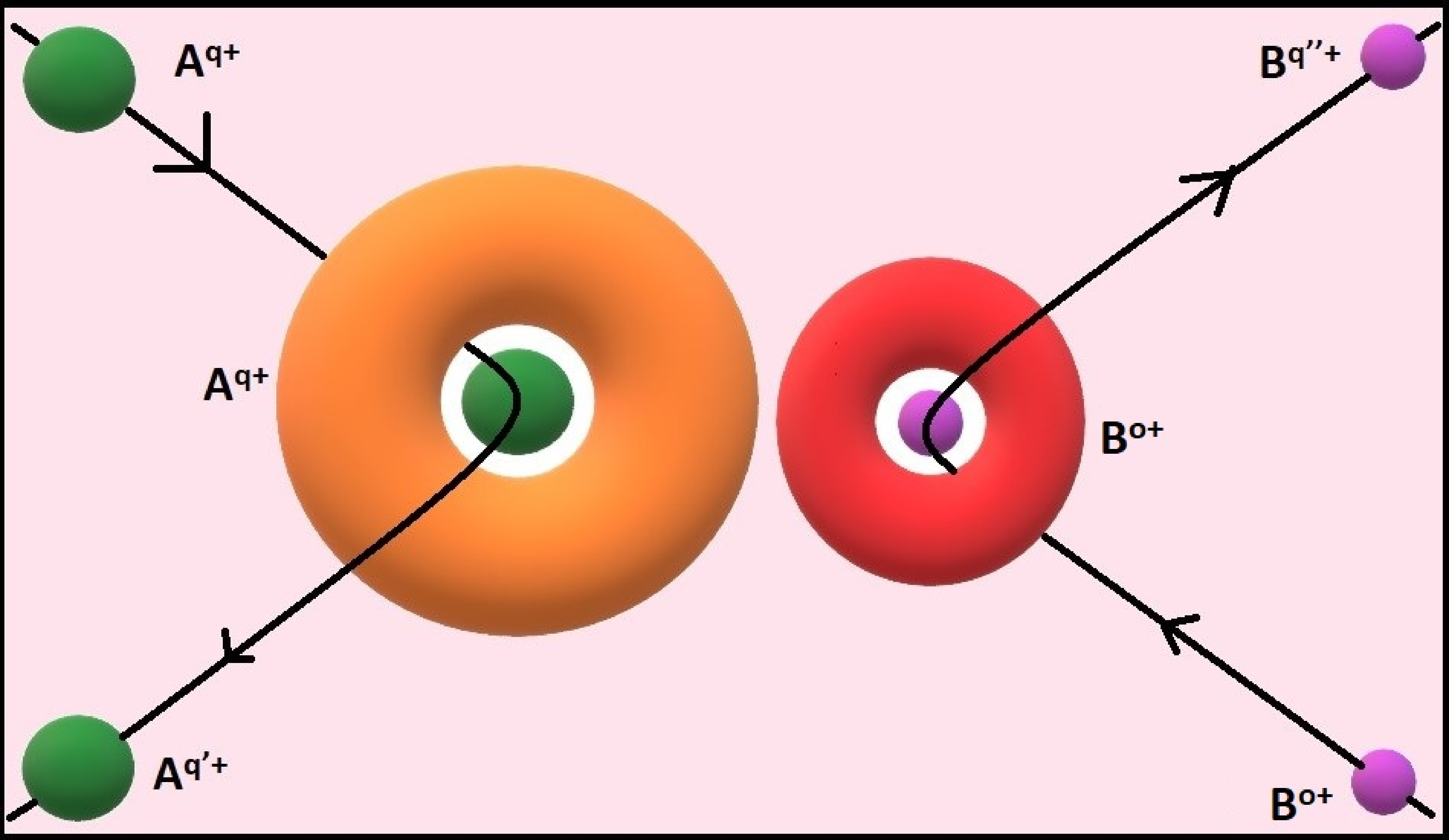}
      \caption{ Schematic of the power radiated due to deceleration by a point charge: The power is radiated in a doughnut about the direction of spontaneous deceleration of the projectile or target nuclei, not in the forward or backward direction. Notice that it is proportional to the square of its deceleration and charge according to the famous Larmor formula.The figure indicates a reaction with inverse kinematics.}
       \label{fig:scattering}
\end{figure}
%%%%%%%%%%%%%%%%%%%%%%%%%%%%%%%%%%%%%%%%%%%%%%%%%%%
\indent {\color{black} Nuclear Coulomb excitation can take place during the heavy-ion collisions at intermediate energy \cite{bertulani2003intermediate} and even at the relativistic energy \cite{wollersheim2011relativistic}.} The origin of nuclear Coulomb excitation is as follows: While the projectile approaching the target nucleus, it faces the nuclear interaction barrier potential \cite{BASS197445} and thus it is retarded, which originates Lienard-Wiechert potential to emit electromagnetic radiation \cite{griffiths11introduction} or nuclear bremsstrahlung. A conceptual schematic diagram is shown in Fig.1. The power is radiated in a doughnut about the direction of spontaneous deceleration of the projectile or target nuclei, not in the forward or backward direction. It can be absorbed by the projectile ion or the target atom and total power radiated ($P$) by the projectile (target) can  be calculated from the famous Larmor formula \cite{griffiths11introduction} in center of mass frame as
\begin{equation}
P = \frac{\mu_0 q^2 f^2}{6\pi c} %\hspace{10}\texbox{and}\hspace{10}    \label{Poynting}
\end{equation} 
\noindent here $\mu_0$ the permeability in vacuum, $q$ the nuclear charge of the projectile (target), $c$ the velocity of light, and $f$ the deceleration of the projectile or target is obtained by the change of velocity of the projectile or target occurring in collision time  $\tau_{coll}$. Two limits of $\tau_{coll}$ are
\begin{equation}
\tau^{max}_{Coll} = \frac{a}{v} ~~~\text{and} ~~~ \tau^{min}_{Coll}=\frac{a\sin{(\theta_{cm}/2)}}{v}.
\label{Coll_time}
\end{equation} 
\noindent here the symmetrized parameter $ a = Z_1Z_2e^2/(m_0 v)$, $m_0$ is the reduced mass, $v$ is projectile ion velocity, and $\theta_{cm}$ is the grazing angle. Note that $\tau^{max}_{Coll}$ causes the minimum power and $\tau^{min}_{Coll}$ the maximum as $f$ varies with inverse squared of the collision time. Since $\theta_{cm}$ is close to 180$^\circ$ at $\leq$ fusion barrier, we get $\tau^{min}_{Coll}=\tau^{max}_{Coll}$ at the sub-barrier energies. This condition holds good for the resonance energy also, which is quite lower than the fusion barrier energy. Resonance  energy, interaction barrier, change of velocity of the projectile ion due to its encounter with the interaction barrier, $\tau^{max}_{Coll}$, and total energy radiated ($P\times\tau^{max}_{Coll}$) in the projectile system are listed in Table I. \\%Note that total energy radiated will be absorbed in both the projectile and target system in the centre of mass frame. \\
%for the reactions $^{56}$Fe + $^{12}$C, $^{58}$Ni + $^{12}$C and $^{63}$Cu + $^{12}$C \cite{Prashant-PRL-2017} %Note that energy is being absorbed in the projectile environment in a very small timescale.\\
\begin{table}
\caption{Total power radiated by the projectile ion due to its retardation by interaction barrier.
Reactions are specified by projectile (Proj.), target (Targ.),  resonance lab energy E$_{res}$ in MeV, and interaction barrier B$_{int}$ in MeV. We list retarded energy of the projectile $E_{ret}$ in MeV, change of velocity $\Delta {v}$ (m/s) occurring in collision time $\tau^{max}_{Coll}$ ($zs$, total energy radiated E$_{rad}$ (keV) in this duration and energy absorbed E$_{abs}$ in the projectile system (keV).}
\begin{tabular}{|c|c|c|c|c|c|c|c|c|}
\hline
Proj. & Targ. & E$_{res}$& B$_{int}$&E$_{ret}$&$\Delta v$& $\tau^{max}_{Coll}$& E$_{rad}$\\
  &  & (MeV)& (MeV)&(MeV)&(m/s)& (zs) &
 (keV)\\\hline
 %\begin{tabular}[c]{@{}c@{}}E$_{abs}$\\(keV)\end{tabular}\\ \hline
 
$^{56}Fe$    & $^{12}C$ & 120 & 98.8  & 21.2  & 7.6 & 2.83 &49.6\\\hline 
$^{58}Ni$    & $^{12}C$ & 134 & 113.8 & 20.4  & 8.3  & 2.50 & 56.2\\\hline 
$^{63}Cu$    & $^{12}C$ & 143 & 125.3 & 17.7  & 9.2 & 2.63 &96.4\\\hline 
\end{tabular}
\end{table}
\begin{figure}[!h]
  \centering
  \includegraphics[width=11cm, height=10cm]{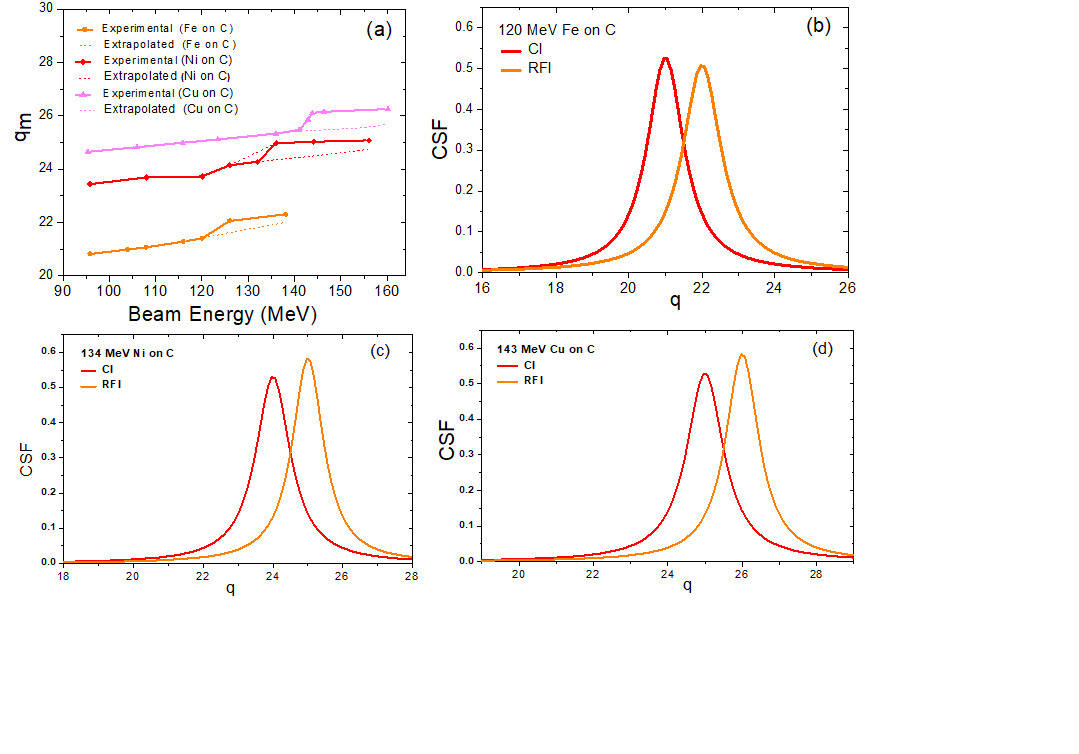}
  \caption{Radiation field ionization (RFI) versus Coulomb ionization (CI): (a) The symbols guided by the solid line are the measured variation of q$_m$ with beam energy and the dotted lines are the same variation but in absence of the nuclear influence \cite{Prashant-PRL-2017}, (b) Charge state distribution (CSD) due CI and RFI for Fe beam on C target, (c)  CSD due CI and RFI for Ni beam on C target, and (d) CSD due CI and RFI for Cu beam on C target. }
  \label{fig:CSD}
\end{figure}
\indent We can notice in Table I that the energy radiated is quite high. Let us see what degree of ionization  can be caused from this energy in the projectile and target systems. The ionization energies vary with the ionic state and these data are readily available in NIST data base \cite{NIST_ASD}. The charge states used for Fe beam of 120 MeV, Ni beam of 134 MeV, and Cu beam of 143 MeV were 9$^+$, 10$^+$, and 11$^+$, respectively \cite{sharma2016experimental}. The radiation energy in every system can ionize it up to the bare ionic stage if we consider the full energy is absorbed in the system. If that is not the case, still high ionic state is achievable because of the fact that the s-electron is ionized faster than the p-electron \cite{schultze}. The absorbed energy can create 1s and 2s shell fully ionized before the p-electrons and the K x-ray spectra can exhibit up to H-like lines. Exactly this fact is experimentally seen as shown in Fig. 2. Fig. 2(a) displays the variation of mean charge states (q$_m$) with beam energies. We can notice that the q$_m$ vary quite smoothly till the resonance occurs at a certain beam energy. This trend signifies that till the resonance point only the Coulomb ionization (CI) is responsible but from this point on wards the nuclear bremsstrahlung being into action. If we fit the q$_m$ data up to the resonance energy with a line and extrapolate it to the higher energies, we get an idea how the q$_m$ would have varied in the post resonance regime in absence of the nuclear bremsstrahlung. This nature is shown by the dotted line in Fig. 2(a). Having gotten the q$_m$ and the fact that the charge state distribution (CSD) inside the foil follows the Lorentzian distribution \cite{sharma2016experimental}, we can numerically obtain the CSD, i.e., a plot of charge state fraction  $F(q)$ versus the charge state $q$, due to the CI as well as radiation field ionization (RFI)  at the resonance energy as follows
\begin{equation}
F(q)=\frac{1}{\pi}\frac{\frac{\Gamma}{2}}{(q-q_m)^2-(\frac{\Gamma}{2})^2}  \label{CSF}
\end{equation}
\noindent
Here distribution width $\Gamma$ is taken from Novikov and Teplova \cite{novikov2014methods}, as follows
%\begin{widetext}
\begin{equation}
\centering
\Gamma(x)= C[1-exp(-(x)^\alpha)][1-exp(-(1-x)^\beta)]\label{Gamma}
\end{equation}
\noindent where $x=q_m/z$, $\alpha=0.23$, $\beta=0.32$, and $C=2.669-0.0098× Z_2+0.058×Z_1+0.00048×Z_1× Z_2$.
Where
\begin{equation}
\Gamma^{2}=[1-(\frac{q_m}{Z})^\frac{5}{3}]\frac{q_m}{4} ~~~\text{and}~~~ \sum_q F(q)=1. \label{Gamma}
\end{equation}
Here, the width $\Gamma$ is taken from the work of \citet{novikov2014methods}. Note that $q_m$ being different for CI and RFI. The CSDs so obtained are shown in Fig. 2(b-d). One can see clearly that the higher charge states are produced due to the RFI from resonance energy on wards. Though both the RFI and CI are the manifestation of the electromagnetic interaction, but they occur through the photons and charge particles, respectively. It implies that the RFI is faster than the CI. If RFI causes the higher charge state in advance, the CI cannot have any role to play on this charge changing process. Thus, the two processes are never in action together. \\ 
\indent To know the difference between the nuclear and atomic timescales for the systems considered, we estimate the characteristic time ($t_0$) for the atomic states from the ratio of the expectation values of electronic radius ($\langle{r}\rangle$) and velocity ($\langle{v}\rangle$) of the corresponding state and it turns out to be in attoseconds. According to the measured x-ray spectra \cite{Prashant-PRL-2017}, on the average the exit channel occurs with the Li-like ions (mean charge state). Considering Li-like $1s2s2p$~$^{2,4}P^o_{1/2,3/2,5/2}$ levels in Fe, Ni, and Cu are mostly populated at the resonance energies. To evaluate $t_0$ for these levels, we have computed $\langle{r}\rangle$ and $\langle{v}\rangle$ by the multi-configuration Dirac-Fock formalism using GRASP2K \cite{jonsson2007grasp2k}. The value of $t_0$ for the three systems mentioned in Table I is found to be 0.383, 0.328, and 0.306 $as$, respectively, which are at least two orders of magnitude larger than the nuclear collision times ($zs$ timescale) when the RFI is taking place. It thus raises a fundamental question: how does then the RFI transcend into the atomic regime ($as$ timescale)?\\%This fact is well explained below in terms of a theoretical calculation. \\
\begin{figure}[!h]
  \centering
    \includegraphics[width=0.9\linewidth]{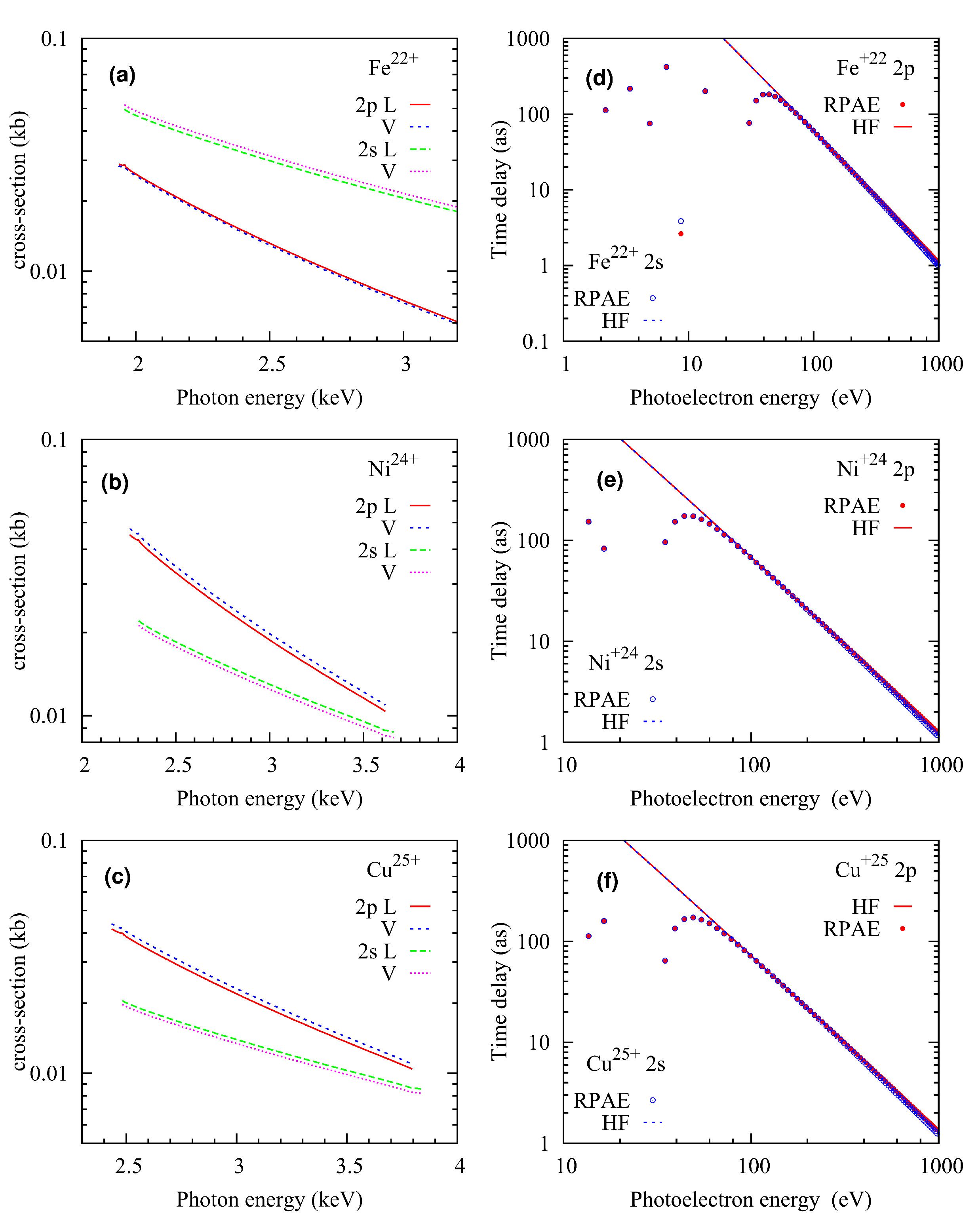}
  \caption{Photoionization cross-section as a function of photon energy of the emitted photons during the heavy-ion collisions shown in the left panel ((a), (b), and (c)).  The EWS time delay due to electromagnetic radiation induced photoionization versus photoelectron energy is shown in the right panel ((d), (e), and (f)). }
\label{timedelay}
\end{figure} 
\indent The above mentioned radiative power may cause both the excitation and ionization in the ions in the short nuclear collision timescales ($zs$). The ionization will give rise to free electrons leaving the ion in ground state, whereas the excitation can lead to autoionization in the multielectron atomic systems. This process may result in excited state that relaxes in to ground state by emitting with higher x-ray energy, as shown in Fig. \ref{fig:CSD}. The autoionization process does not occur instantly as the electron has to move from the interaction regime to an interaction free regime, which introduces a time delay because of the difference between the density of states of the two regions \cite{ahmed2004number}. The energy-integral of time delay is an adiabatic invariant in quantum scattering theory and it provides a quantization condition for resonances \cite{jain2005}. The delay in the autoionization process, called as the Eisenbud-Wigner-Smith (EWS) time delay \cite{wigner55,smith60}, has been measured in a recent experiment \cite{schultze}. To estimate the EWS time delay as well as the photoionization cross-section, we have employed the nonrelativistic versions of the random-phase approximation with exchange (RPAE) and Hartree-Fock (HF) methods \cite{ASK2015}. The partial photoionization cross-section for the transition from an occupied state $n_il_i$ to the photoelectron continuum state $kl$ is calculated as
\begin{equation}
\label{CS-HF} \sigma_{n_il_i\to kl}(\omega) = \frac43 \pi^2\alpha a_0^2\omega
\left| \langle k l\,\|\,\hat D\,\|n_il_i \rangle \right|^2 \,  
\end{equation} 
\noindent where $\omega$ being the photon energy, $\alpha$ the fine structure constant
and $a_0$ the Bohr radius are used in atomic units $e=m=\hbar=1$.
In the independent electron Hartree-Fock (HF) approximation, the
reduced dipole matrix element is evaluated as a radial integral given below
\be
\label{reduced} \langle kl \|\,\hat D\,\|n_il_i\rangle=[l][l_i] \left(
\begin{array}{rrr} l&1&l_i\\ 0&0&0\\
\end{array}\right)\\\int r^2dr\, R_{kl}(r)\,r\,R_{n_il_i}(r)\,
\ee
where the notation $[l] = \sqrt{2l+1}$ is used. The basis of occupied atomic states $\|n_il_i\rangle$ is defined by the self-consistent HF method and calculated using the computer code
\cite{CCR76}. The continuum electron orbitals $\langle kl \|$ are defined
within the frozen-core HF approximation and evaluated using the
computer code \cite{CCR79}. \\
\indent In the RPAE, the reduced dipole matrix element is found by summing an infinite sequence of Coulomb interactions between the photoelectron and the hole in the ionized shell. A virtual
excitation in the shell $j$ to the ionized electron state $\k'$ may
affect the final ionization channel from the shell $i$. This way RPAE
accounts for the effect of inter-shell $i\leftrightarrow j$
correlation, also known as inter-channel coupling.  It is important to
note that, within the RPAE framework, the reduced dipole matrix element is 
complex and, thereby, adds to the phase of the dipole amplitude.\\
\indent The photoelectron group delay, which is the energy derivative of the
phase of the complex photoionization amplitude, is evaluated as
\begin{equation}
\label{delay}
\tau = {d\over dE} \arg f(E)\equiv
{\rm Im} \Big[ f'(E)/f(E) \Big].
\end{equation}
\noindent Here $f(E)$ is used as a shortcut for the amplitude 
$\langle\psi^{(-)}_\k|\hat z|\phi_i\rangle$ evaluated
for $E=k^2/2$ and $\hat\k\parallel\z$, where $\psi^{(-)}_\k$ is an incoming scattering
state with the given photoelectron momentum $\k$ \cite{A90}.\\
\indent For the photoionization and time delay calculations, we have considered the charge species Fe$^{22+}$, Ni$^{24+}$, and Cu$^{25+}$ for the systems $^{56}$Fe + $^{12}$C, $^{58}$Ni + $^{12}$C, and $^{63}$Cu + $^{12}$C, respectively, as the Li-like ions are observed \cite{Prashant-PLA} after the autoionizaion. It is found that the EWS time delay, $\tau$, is proportional to  $\varepsilon ^{- 3/2} ln (1/\varepsilon$), where $\varepsilon$ is the photon energy and the elastic scattering phase, $\sigma$ ($\approx$ $\eta$ $ln {|\eta|}$, $\eta = -Z/\sqrt{2\varepsilon}$) are divergent near the threshold because of the Coulomb singularity \cite{ASK2015}. To remove this singularity, we have cut off the low energy photoelectrons for time delay calculations in present computations. The results of the computations are presented in  Fig.\ref{timedelay} and the  photoionization cross-sections are very close for the length ({\textit{L}}) and velocity ({\textit{V}}) gauges. The photoionization cross-sections for 2p electron in Ni$^{24+}$ and Cu$^{25+}$ are higher than that of 2s electron for all photon energies. Whereas, for Fe$^{22+}$, the  photoionization cross-sections for 2s electron is higher than the 2p electron. \\
%Further, the Wigner-Smith time delay  can transfer the  \rrt{orbiting  induced ionization } triggered  in the nuclear timescale ($zs$) to a few hundreds of $as$ for all the three systems considered here. Thereby, the photoionization phenomenon can occur in the atomic timescale and one measures the x-ray emissions. \\
%%%%%%%%%%
\indent  Dissociation of the excited state can only occur in $as$ or greater timescale because no atomic states exist with lifetime shorter than an $as$. This sort of occurrence is due to the fact that the emission of an electron from an excited state is delayed by $as$ due to electron correlation effects \cite{de2002time}. For example, photoemission of an electron is delayed up to six attoseconds \cite{ossiander2017attosecond}. Laser induced photoionization of different states in Ne are delayed differently. The 2p state is 21$\pm5$ $as$ more delayed than the 2s state in Ne because of multi electron-correlation dynamics \cite{schultze}. This picture has been theoretically reproduced  using above mentioned approach and the same is applied for three test cases. \\
\begin{figure}
 \vspace{3mm}
  \centering
   \includegraphics[width=9cm, height=10cm]{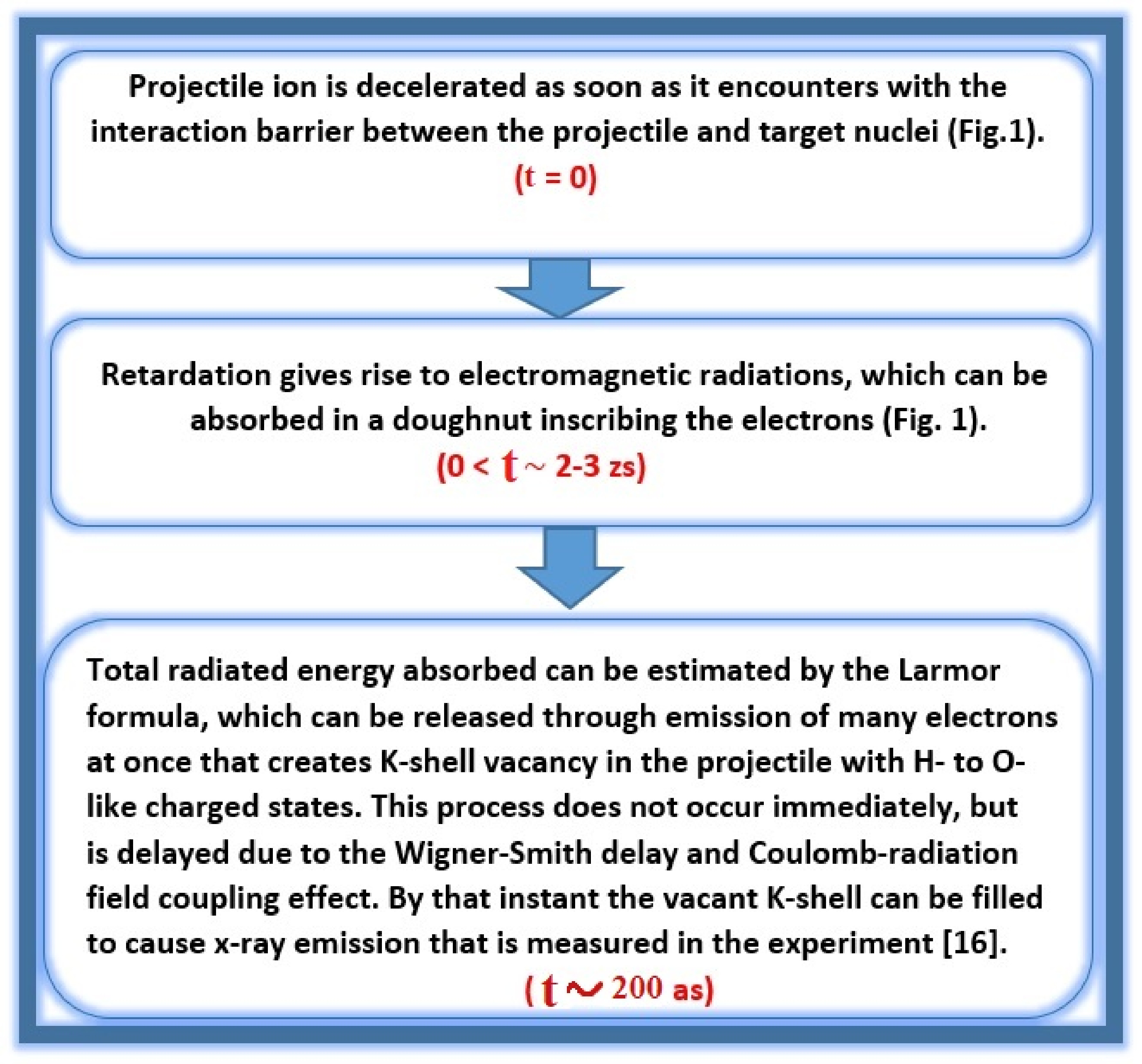}  \caption{Schematic of the overall x-ray emission mechanism at the resonance energies \cite{Prashant-PRL-2017} and beyond, where the atomic process is influenced by the nuclear bremsstrahlung process due to heavy-ion collisions. See the text for details.}
    \label{fig:Schematic}
\end{figure}
\indent Let us consider the RFI may result in Be-like ions and at least two electrons are in excited states. Naturally, the autoionization process of such Be-like ions will lead to the Li-like ions through a certain delay as discussed above.  This delay can be calculated as a function of photo-electron energy using the Eqn. \ref{delay} as shown in Fig.\ref{timedelay}. One can see that the delays are in $as$ range for every photoelectron energy and the maximum delay is about 200 $as$ for the photoelectrons $\approx$ 80 eV.\\
\indent Effect of nuclear bremsstrahlung in the projectile ion beyond the resonance energies is schematically described in Fig.\ref{fig:Schematic}. Projectile ion can be retarded as soon as it encounters with interaction barrier ($t=0$). Retardation is maximum at the saddle point, which gives rise to electromagnetic radiation from the projectile ion (target atom) that can be absorbed in a doughnut containing the projectile (target) electrons. This incident occurs during the short collision time in $zs$ duration as given in Table I. Total radiated energy absorbed as estimated by the Larmor formula is quite large that can  be released through emission of many electrons at once (because a bunch of electrons got sufficient energy to be ionized and electron-electron correlation does not allow to an individual electron to escape and leaving the projectile ion in much higher charged states (H- to O-like ions according to Fig. \ref{fig:CSD})). This photon induced ionization can take place at a duration larger (up to $200 as$) than the atomic characteristic times ($< 1~as$) so that Blair and Anholt \cite{blair1982theory} criteria mentioned above can be fulfilled. At this condition  both the photoionization and photon deexcitation due to filling of vacant K-shell can be possible. Accordingly one can measure the phenomenon in x-ray spectroscopy experiments \cite{Prashant-PRL-2017}. \\ 
\indent Though zeptosecond laser is far from reality till date but the present work provides a new opportunity for atomic physics research at the nuclear sub- and over-barrier regions, where a laser can be replaced by the radiation field produced by the deceleration of the projectile (target) at the Coulomb interaction barrier. Since, at the saddle point the radiated power is maximum, it can produce the radiations of a fixed frequency. Using this monochromatic source we can study the atomic events well as atomic phenomena are longer than $as$ and the source is as fast as a few $zs$. \\
\indent To conclude, nuclear bremsstrahlung induced radiation energy absorbed in the electronic environment of the projectile ions leading to highly charged ions in the multiply excited states due to ionization and excitation in a short nuclear $zs$ timescales, which results in autoionization. The EWS time delay associated to such autoionization is about 200 $as$ (atomic timescale) for three representative cases as used in the experiments \cite{Prashant-PRL-2017}. Hence, we revealed that the events in nuclear timescales can cause the phenomenon in atomic timescales. Significantly, present work provides a new opportunity for atomic physics research at the nuclear sub-barrier regions, where the laser is being replaced by the nuclear bremsstrahlung.
\bibliographystyle{elsarticle-num-names}
\bibliography{amain.bbl}

\begin{thebibliography}{36}
\expandafter\ifx\csname natexlab\endcsname\relax\def\natexlab#1{#1}\fi
\providecommand{\url}[1]{\texttt{#1}}
\providecommand{\href}[2]{#2}
\providecommand{\path}[1]{#1}
\providecommand{\DOIprefix}{doi:}
\providecommand{\ArXivprefix}{arXiv:}
\providecommand{\URLprefix}{URL: }
\providecommand{\Pubmedprefix}{pmid:}
\providecommand{\doi}[1]{\href{http://dx.doi.org/#1}{\path{#1}}}
\providecommand{\Pubmed}[1]{\href{pmid:#1}{\path{#1}}}
\providecommand{\bibinfo}[2]{#2}
\ifx\xfnm\relax \def\xfnm[#1]{\unskip,\space#1}\fi
%Type = Article
\bibitem[{Basbas et~al.(1973)Basbas, Brandt, and Laubert}]{basbas1974universal}
\bibinfo{author}{G.~Basbas}, \bibinfo{author}{W.~Brandt},
  \bibinfo{author}{R.~Laubert},
\newblock \bibinfo{title}{Universal cross sections for k-shell ionization by
  heavy charged particles. i. low particle velocities},
\newblock \bibinfo{journal}{Physical Review A} \bibinfo{volume}{7}
  (\bibinfo{year}{1973}) \bibinfo{pages}{983}.
%Type = Article
\bibitem[{Brandt and Lapicki(1974)}]{brandt1974binding}
\bibinfo{author}{W.~Brandt}, \bibinfo{author}{G.~Lapicki},
\newblock \bibinfo{title}{Binding and coulomb-deflection effects in l-shell
  coulomb ionization by heavy charged particles. low particle velocities},
\newblock \bibinfo{journal}{Physical Review A} \bibinfo{volume}{10}
  (\bibinfo{year}{1974}) \bibinfo{pages}{474}.
%Type = Article
\bibitem[{Brandt and Lapicki(1979)}]{brandt1979shell}
\bibinfo{author}{W.~Brandt}, \bibinfo{author}{G.~Lapicki},
\newblock \bibinfo{title}{L-shell coulomb ionization by heavy charged
  particles},
\newblock \bibinfo{journal}{Physical Review A} \bibinfo{volume}{20}
  (\bibinfo{year}{1979}) \bibinfo{pages}{465}.
%Type = Article
\bibitem[{Barber and Helm(1952)}]{barber1952evidence}
\bibinfo{author}{W.~Barber}, \bibinfo{author}{R.~Helm},
\newblock \bibinfo{title}{Evidence for k-shell ionization accompanying the
  alpha-decay of po 210},
\newblock \bibinfo{journal}{Physical Review} \bibinfo{volume}{86}
  (\bibinfo{year}{1952}) \bibinfo{pages}{275}.
%Type = Article
\bibitem[{Ciocchetti and Molinari(1965)}]{ciocchetti1965k}
\bibinfo{author}{G.~Ciocchetti}, \bibinfo{author}{A.~Molinari},
\newblock \bibinfo{title}{K electron shell ionization and nuclear reactions},
\newblock \bibinfo{journal}{Il Nuovo Cimento B} \bibinfo{volume}{40}
  (\bibinfo{year}{1965}) \bibinfo{pages}{69--86}.
%Type = Article
\bibitem[{Blair et~al.(1978)Blair, Dyer, Snover, and
  Trainor}]{blair1978nuclear}
\bibinfo{author}{J.~Blair}, \bibinfo{author}{P.~Dyer},
  \bibinfo{author}{K.~Snover}, \bibinfo{author}{T.~Trainor},
\newblock \bibinfo{title}{Nuclear "time-delay" and x-ray-proton coincidences
  near a nuclear scattering resonance},
\newblock \bibinfo{journal}{Physical Review Letters} \bibinfo{volume}{41}
  (\bibinfo{year}{1978}) \bibinfo{pages}{1712}.
%Type = Article
\bibitem[{Duinker et~al.(1980)Duinker, Van~Eck, and
  Niehaus}]{duinker1980experimental}
\bibinfo{author}{W.~Duinker}, \bibinfo{author}{J.~Van~Eck},
  \bibinfo{author}{A.~Niehaus},
\newblock \bibinfo{title}{Experimental evidence for the influence of
  inner-shell ionization on resonant nuclear scattering},
\newblock \bibinfo{journal}{Physical Review Letters} \bibinfo{volume}{45}
  (\bibinfo{year}{1980}) \bibinfo{pages}{2102}.
%Type = Article
\bibitem[{Chemin et~al.(1981)Chemin, Anholt, Stoller, Meyerhot, and
  Amundsen}]{chemin1981measurement}
\bibinfo{author}{J.~Chemin}, \bibinfo{author}{R.~Anholt},
  \bibinfo{author}{C.~Stoller}, \bibinfo{author}{W.~Meyerhot},
  \bibinfo{author}{P.~Amundsen},
\newblock \bibinfo{title}{Measurement of sr 88 k-shell ionization probability
  across the nuclear elastic-scattering resonance at 5060 kev},
\newblock \bibinfo{journal}{Physical Review A} \bibinfo{volume}{24}
  (\bibinfo{year}{1981}) \bibinfo{pages}{1218}.
%Type = Article
\bibitem[{Blair and Anholt(1982)}]{blair1982theory}
\bibinfo{author}{J.~Blair}, \bibinfo{author}{R.~Anholt},
\newblock \bibinfo{title}{Theory of k-shell ionization during nuclear resonance
  scattering},
\newblock \bibinfo{journal}{Physical Review A} \bibinfo{volume}{25}
  (\bibinfo{year}{1982}) \bibinfo{pages}{907}.
%Type = Article
\bibitem[{Greenberg et~al.(1977)Greenberg, Bokemeyer, Emling, Grosse, Schwalm,
  and Bosch}]{greenberg1977impact}
\bibinfo{author}{J.~Greenberg}, \bibinfo{author}{H.~Bokemeyer},
  \bibinfo{author}{H.~Emling}, \bibinfo{author}{E.~Grosse},
  \bibinfo{author}{D.~Schwalm}, \bibinfo{author}{F.~Bosch},
\newblock \bibinfo{title}{Impact-parameter dependence of vacancy production in
  strongly bound quasimolecular states of heavy collision systems},
\newblock \bibinfo{journal}{Physical Review Letters} \bibinfo{volume}{39}
  (\bibinfo{year}{1977}) \bibinfo{pages}{1404}.
%Type = Article
\bibitem[{Betz et~al.(1976)Betz, Soff, M{\"u}ller, and
  Greiner}]{betz1976direct}
\bibinfo{author}{W.~Betz}, \bibinfo{author}{G.~Soff},
  \bibinfo{author}{B.~M{\"u}ller}, \bibinfo{author}{W.~Greiner},
\newblock \bibinfo{title}{Direct formation of quasimolecular 1 s $\sigma$
  vacancies in uranium-uranium collisions},
\newblock \bibinfo{journal}{Physical Review Letters} \bibinfo{volume}{37}
  (\bibinfo{year}{1976}) \bibinfo{pages}{1046}.
%Type = Article
\bibitem[{Sharma and Nandi(2017)}]{Prashant-PRL-2017}
\bibinfo{author}{P.~Sharma}, \bibinfo{author}{T.~Nandi},
\newblock \bibinfo{title}{Shakeoff ionization near the coulomb barrier energy},
\newblock \bibinfo{journal}{Phys. Rev. Lett.} \bibinfo{volume}{119}
  (\bibinfo{year}{2017}) \bibinfo{pages}{203401}. \URLprefix
  \url{https://link.aps.org/doi/10.1103/PhysRevLett.119.203401}.
  \DOIprefix\doi{10.1103/PhysRevLett.119.203401}.
%Type = Article
\bibitem[{Sharma and Nandi(2016)}]{Prashant-PLA}
\bibinfo{author}{P.~Sharma}, \bibinfo{author}{T.~Nandi},
\newblock \bibinfo{title}{X-ray spectroscopy: An experimental technique to
  measure charge state distribution during ion–solid interaction},
\newblock \bibinfo{journal}{Physics Letters A} \bibinfo{volume}{380}
  (\bibinfo{year}{2016}) \bibinfo{pages}{182 -- 187}. \URLprefix
  \url{http://www.sciencedirect.com/science/article/pii/S037596011500818X}.
  \DOIprefix\doi{https://doi.org/10.1016/j.physleta.2015.09.031}.
%Type = Article
\bibitem[{Jakuba{\ss}a and Kleber(1975)}]{jakubassa1975bremsstrahlung}
\bibinfo{author}{D.~H. Jakuba{\ss}a}, \bibinfo{author}{M.~Kleber},
\newblock \bibinfo{title}{Bremsstrahlung in heavy-ion reactions},
\newblock \bibinfo{journal}{Zeitschrift f{\"u}r Physik A Atoms and Nuclei}
  \bibinfo{volume}{273} (\bibinfo{year}{1975}) \bibinfo{pages}{29--35}.
%Type = Article
\bibitem[{Griffy and Biedenharn(1962)}]{griffy1962nuclear}
\bibinfo{author}{T.~Griffy}, \bibinfo{author}{L.~Biedenharn},
\newblock \bibinfo{title}{Nuclear resonance effects in coulomb excitation},
\newblock \bibinfo{journal}{Nuclear Physics} \bibinfo{volume}{32}
  (\bibinfo{year}{1962}) \bibinfo{pages}{273--285}.
%Type = Article
\bibitem[{Alder et~al.(1956)Alder, Bohr, Huus, Mottelson, and Winther}]{Alder}
\bibinfo{author}{K.~Alder}, \bibinfo{author}{A.~Bohr},
  \bibinfo{author}{T.~Huus}, \bibinfo{author}{B.~Mottelson},
  \bibinfo{author}{A.~Winther},
\newblock \bibinfo{title}{{Study of Nuclear Structure by Electromagnetic
  Excitation with Accelerated Ions}},
\newblock \bibinfo{journal}{Reviews of Modern Physics} \bibinfo{volume}{28}
  (\bibinfo{year}{1956}) \bibinfo{pages}{432--542}. \URLprefix
  \url{http://link.aps.org/doi/10.1103/RevModPhys.28.432}.
%Type = Article
\bibitem[{Izumoto et~al.(1981)Izumoto, Lui, Youngblood, Udagawa, and
  Tamura}]{izumoto1981coulomb}
\bibinfo{author}{T.~Izumoto}, \bibinfo{author}{Y.-W. Lui},
  \bibinfo{author}{D.~H. Youngblood}, \bibinfo{author}{T.~Udagawa},
  \bibinfo{author}{T.~Tamura},
\newblock \bibinfo{title}{Coulomb excitation of the giant dipole resonance in
  light-ion inelastic scattering from pb 208},
\newblock \bibinfo{journal}{Physical Review C} \bibinfo{volume}{24}
  (\bibinfo{year}{1981}) \bibinfo{pages}{2179}.
%Type = Article
\bibitem[{Bertulani et~al.(2003)Bertulani, Stuchbery, Mertzimekis, and
  Davies}]{bertulani2003intermediate}
\bibinfo{author}{C.~Bertulani}, \bibinfo{author}{A.~Stuchbery},
  \bibinfo{author}{T.~Mertzimekis}, \bibinfo{author}{A.~Davies},
\newblock \bibinfo{title}{Intermediate energy coulomb excitation as a probe of
  nuclear structure at radioactive beam facilities},
\newblock \bibinfo{journal}{Physical Review C} \bibinfo{volume}{68}
  (\bibinfo{year}{2003}) \bibinfo{pages}{044609}.
%Type = Article
\bibitem[{Wollersheim(2011)}]{wollersheim2011relativistic}
\bibinfo{author}{H.-J. Wollersheim},
\newblock \bibinfo{title}{Relativistic coulomb excitation: From rising to
  prespec.},
\newblock \bibinfo{journal}{Acta Physica Polonica B} \bibinfo{volume}{42}
  (\bibinfo{year}{2011}).
%Type = Article
\bibitem[{Bass(1974)}]{BASS197445}
\bibinfo{author}{R.~Bass},
\newblock \bibinfo{title}{Fusion of heavy nuclei in a classical model},
\newblock \bibinfo{journal}{Nuclear Physics A} \bibinfo{volume}{231}
  (\bibinfo{year}{1974}) \bibinfo{pages}{45 -- 63}. \URLprefix
  \url{http://www.sciencedirect.com/science/article/pii/0375947474902929}.
  \DOIprefix\doi{https://doi.org/10.1016/0375-9474(74)90292-9}.
%Type = Article
\bibitem[{Griffiths(????)}]{griffiths11introduction}
\bibinfo{author}{D.~J. Griffiths},
\newblock \bibinfo{title}{introduction to electrodynamics, 3rd-ed, 2007},
\newblock \bibinfo{journal}{published by Dorling Kindersely Pvt. Ltd, NO-11,
  community center, Punchsheel park, New Delhi-110017, India}  (????)
  \bibinfo{pages}{521--522}.
%Type = Misc
\bibitem[{Kramida et~al.(2019)Kramida, {Yu.~Ralchenko}, Reader, and {and NIST
  ASD Team}}]{NIST_ASD}
\bibinfo{author}{A.~Kramida}, \bibinfo{author}{{Yu.~Ralchenko}},
  \bibinfo{author}{J.~Reader}, \bibinfo{author}{{and NIST ASD Team}},
  \bibinfo{howpublished}{{NIST Atomic Spectra Database (ver. 5.7.1), [Online].
  Available: {\tt{https://physics.nist.gov/asd}} [2020, August 22]. National
  Institute of Standards and Technology, Gaithersburg, MD.}},
  \bibinfo{year}{2019}.
%Type = Article
\bibitem[{Sharma and Nandi(2016)}]{sharma2016experimental}
\bibinfo{author}{P.~Sharma}, \bibinfo{author}{T.~Nandi},
\newblock \bibinfo{title}{Experimental evidence of beam-foil plasma creation
  during ion-solid interaction},
\newblock \bibinfo{journal}{Physics of Plasmas} \bibinfo{volume}{23}
  (\bibinfo{year}{2016}) \bibinfo{pages}{083102}.
%Type = Article
\bibitem[{Schultze et~al.(2010)Schultze, Fie{\ss}, Karpowicz, Gagnon, Korbman,
  Hofstetter, Neppl, Cavalieri, Komninos, Mercouris et~al.}]{schultze}
\bibinfo{author}{M.~Schultze}, \bibinfo{author}{M.~Fie{\ss}},
  \bibinfo{author}{N.~Karpowicz}, \bibinfo{author}{J.~Gagnon},
  \bibinfo{author}{M.~Korbman}, \bibinfo{author}{M.~Hofstetter},
  \bibinfo{author}{S.~Neppl}, \bibinfo{author}{A.~L. Cavalieri},
  \bibinfo{author}{Y.~Komninos}, \bibinfo{author}{T.~Mercouris}, et~al.,
\newblock \bibinfo{title}{Delay in photoemission},
\newblock \bibinfo{journal}{science} \bibinfo{volume}{328}
  (\bibinfo{year}{2010}) \bibinfo{pages}{1658--1662}.
%Type = Article
\bibitem[{Novikov and Teplova(2014)}]{novikov2014methods}
\bibinfo{author}{N.~Novikov}, \bibinfo{author}{Y.~A. Teplova},
\newblock \bibinfo{title}{Methods of estimation of equilibrium charge
  distribution of ions in solid and gaseous media},
\newblock \bibinfo{journal}{Physics Letters A} \bibinfo{volume}{378}
  (\bibinfo{year}{2014}) \bibinfo{pages}{1286--1289}.
%Type = Article
\bibitem[{J{\"o}nsson et~al.(2007)J{\"o}nsson, He, Fischer, and
  Grant}]{jonsson2007grasp2k}
\bibinfo{author}{P.~J{\"o}nsson}, \bibinfo{author}{X.~He},
  \bibinfo{author}{C.~F. Fischer}, \bibinfo{author}{I.~Grant},
\newblock \bibinfo{title}{The grasp2k relativistic atomic structure package},
\newblock \bibinfo{journal}{Computer Physics Communications}
  \bibinfo{volume}{177} (\bibinfo{year}{2007}) \bibinfo{pages}{597--622}.
%Type = Article
\bibitem[{Ahmed and Jain(2004)}]{ahmed2004number}
\bibinfo{author}{Z.~Ahmed}, \bibinfo{author}{S.~R. Jain},
\newblock \bibinfo{title}{Number of quantal resonances},
\newblock \bibinfo{journal}{Journal of Physics A: Mathematical and General}
  \bibinfo{volume}{37} (\bibinfo{year}{2004}) \bibinfo{pages}{867}.
%Type = Article
\bibitem[{Jain(2005)}]{jain2005}
\bibinfo{author}{S.~R. Jain},
\newblock \bibinfo{title}{Resonances and adiabatic invariance in classical and
  quantum scattering theory},
\newblock \bibinfo{journal}{Physics Letters A} \bibinfo{volume}{335}
  (\bibinfo{year}{2005}) \bibinfo{pages}{83 -- 87}. \URLprefix
  \url{http://www.sciencedirect.com/science/article/pii/S0375960104017207}.
  \DOIprefix\doi{https://doi.org/10.1016/j.physleta.2004.12.014}.
%Type = Article
\bibitem[{Wigner(1955)}]{wigner55}
\bibinfo{author}{E.~P. Wigner},
\newblock \bibinfo{title}{Lower limit for the energy derivative of the
  scattering phase shift},
\newblock \bibinfo{journal}{Phys. Rev.} \bibinfo{volume}{98}
  (\bibinfo{year}{1955}) \bibinfo{pages}{145--147}. \URLprefix
  \url{https://link.aps.org/doi/10.1103/PhysRev.98.145}.
  \DOIprefix\doi{10.1103/PhysRev.98.145}.
%Type = Article
\bibitem[{Smith(1960)}]{smith60}
\bibinfo{author}{F.~T. Smith},
\newblock \bibinfo{title}{Lifetime matrix in collision theory},
\newblock \bibinfo{journal}{Phys. Rev.} \bibinfo{volume}{118}
  (\bibinfo{year}{1960}) \bibinfo{pages}{349--356}. \URLprefix
  \url{https://link.aps.org/doi/10.1103/PhysRev.118.349}.
  \DOIprefix\doi{10.1103/PhysRev.118.349}.
%Type = Article
\bibitem[{Kheifets et~al.(2015)Kheifets, Saha, Deshmukh, Keating, and
  Manson}]{ASK2015}
\bibinfo{author}{A.~S. Kheifets}, \bibinfo{author}{S.~Saha},
  \bibinfo{author}{P.~C. Deshmukh}, \bibinfo{author}{D.~A. Keating},
  \bibinfo{author}{S.~T. Manson},
\newblock \bibinfo{title}{Dipole phase and photoelectron group delay in
  inner-shell photoionization},
\newblock \bibinfo{journal}{Phys. Rev. A} \bibinfo{volume}{92}
  (\bibinfo{year}{2015}) \bibinfo{pages}{063422}. \URLprefix
  \url{https://link.aps.org/doi/10.1103/PhysRevA.92.063422}.
  \DOIprefix\doi{10.1103/PhysRevA.92.063422}.
%Type = Article
\bibitem[{Chernysheva et~al.(1976)Chernysheva, Cherepkov, and
  Radojevi{\'c}}]{CCR76}
\bibinfo{author}{L.~Chernysheva}, \bibinfo{author}{N.~Cherepkov},
  \bibinfo{author}{V.~Radojevi{\'c}},
\newblock \bibinfo{title}{Self-consistent field hartree-fock program for
  atoms},
\newblock \bibinfo{journal}{Computer Physics Communications}
  \bibinfo{volume}{11} (\bibinfo{year}{1976}) \bibinfo{pages}{57--73}.
%Type = Article
\bibitem[{Chernysheva et~al.(1979)Chernysheva, Cherepkov, and
  Radojevic}]{CCR79}
\bibinfo{author}{L.~Chernysheva}, \bibinfo{author}{N.~Cherepkov},
  \bibinfo{author}{V.~Radojevic},
\newblock \bibinfo{title}{Frozen core hartree-fock program for atomic discrete
  and continuous states},
\newblock \bibinfo{journal}{Computer Physics Communications}
  \bibinfo{volume}{18} (\bibinfo{year}{1979}) \bibinfo{pages}{87--100}.
%Type = Book
\bibitem[{Amusia(2013)}]{A90}
\bibinfo{author}{M.~Y. Amusia}, \bibinfo{title}{Atomic photoeffect},
  \bibinfo{publisher}{Springer Science \& Business Media},
  \bibinfo{year}{2013}.
%Type = Article
\bibitem[{de~Carvalho and Nussenzveig(2002)}]{de2002time}
\bibinfo{author}{C.~A. de~Carvalho}, \bibinfo{author}{H.~M. Nussenzveig},
\newblock \bibinfo{title}{Time delay},
\newblock \bibinfo{journal}{Physics Reports} \bibinfo{volume}{364}
  (\bibinfo{year}{2002}) \bibinfo{pages}{83--174}.
%Type = Article
\bibitem[{Ossiander et~al.(2017)Ossiander, Siegrist, Shirvanyan, Pazourek,
  Sommer, Latka, Guggenmos, Nagele, Feist, Burgd{\"o}rfer
  et~al.}]{ossiander2017attosecond}
\bibinfo{author}{M.~Ossiander}, \bibinfo{author}{F.~Siegrist},
  \bibinfo{author}{V.~Shirvanyan}, \bibinfo{author}{R.~Pazourek},
  \bibinfo{author}{A.~Sommer}, \bibinfo{author}{T.~Latka},
  \bibinfo{author}{A.~Guggenmos}, \bibinfo{author}{S.~Nagele},
  \bibinfo{author}{J.~Feist}, \bibinfo{author}{J.~Burgd{\"o}rfer}, et~al.,
\newblock \bibinfo{title}{Attosecond correlation dynamics},
\newblock \bibinfo{journal}{Nature Physics} \bibinfo{volume}{13}
  (\bibinfo{year}{2017}) \bibinfo{pages}{280--285}.

\end{thebibliography}
\end{document}